\documentclass[journal]{IEEEtran}
\ifCLASSINFOpdf
\else
\fi
\usepackage{moreverb}
\usepackage{amssymb}
\usepackage{latexsym}
\usepackage{amsfonts}
\usepackage{amsthm}
\usepackage{amsmath}
\usepackage{graphics}
\usepackage{setspace}
\usepackage{graphicx}
\usepackage{epstopdf}
\usepackage{color}
\usepackage{float}
\usepackage{wrapfig}
\usepackage{color}
\usepackage{rotating}
\usepackage{array}
\usepackage{dblfloatfix}
\usepackage[hyphens]{url}
\usepackage[table]{xcolor}
\usepackage{multirow}
\usepackage{graphics}
\usepackage{setspace}
\usepackage{algorithm}
\usepackage{algorithmic}
\usepackage{graphicx}
\usepackage{epstopdf}
\usepackage{color}
\usepackage{enumitem}
\usepackage{float}
\usepackage{wrapfig}
\usepackage{color}
\usepackage{xspace}

\begin{document}
%
\title{A Contextual Investigation of Location in the Home Using Bluetooth Low Energy  Beacons}
%
%
%

\author{Charith Perera,
        Saeed Aghaee,
        Ramsey Faragher,
        Robert Harle,
        and Alan Blackwell
}

%
%

\markboth{Technical Report, March, 2017}%
{}

%



\maketitle

\begin{abstract}
Location sensing is a key enabling technology for Ubicomp to support contextual interaction. However, the laboratories where calibrated testing of location technologies is done are very different to the domestic situations where `context' is a problematic social construct. This study reports measurements of Bluetooth beacons, informed by laboratory studies, but done in diverse domestic settings. The design of these surveys has been motivated by the natural environment implied in the Bluetooth beacon standards – relating the technical environment of the beacon to the function of spaces within the home. This research method can be considered as a situated, `ethnographic' technical response to the study of physical infrastructure that arises through social processes. The results offer insights for the future design of `seamful' approaches to indoor location sensing, and to the ways that context might be constructed and interpreted in a seamful manner.
\end{abstract}

\begin{IEEEkeywords}
Location,  Beacons
\end{IEEEkeywords}

%
\IEEEpeerreviewmaketitle

\section{Introduction}

\IEEEPARstart{M}{any} Ubicomp services rely on a model of context in order to interpret user actions and needs. However, a classic paper by Dourish \cite{Dourish2004} challenged the way that context models are derived only from sensor and activity data, while failing to recognise the nature of human interaction. Dourish’s main contribution was to note the ways that context is jointly established in a kind of conversation, rather than simply being delivered as a technical product feature. In this research, we explore some technical implications of that perspective on context sensing. 
Location technologies for Ubicomp represent an important and growing element of context. One view of location sensing is that it offers a reference grid – a spatial map on which the user and relevant world features are marked. Outdoor positioning products based on GPS often present their data in precisely this way. For some years, developers have been working toward indoor positioning properties that could do the same in relation to a spatial map of a given building. However, it is possible to take an alternative approach to location, for example as expressed in Chalmers et al \cite{Chalmers2003} proposal of ``seamful design'' that acknowledges the gaps and inaccuracies in GPS signal coverage, using them as a design resource rather than a system failing.

In this paper, we apply these perspectives – the human understanding of context, and the strategy of seamful design – to the indoor location technology of Bluetooth low energy beacons. There have been substantial advances in the calibration of these beacons as a basis for establishing an accurate positioning grid in controlled conditions \cite{Faragher2015}. However, current commercial applications do not currently emphasise grid position, but simply trigger services based on proximity to the beacon. In a recent project, we carried out a design exercise in which the ``seamful'' approach was applied in a museum context to deal with the ambiguity resulting from the very large numbers of objects in a museum, that are too close together to reliably be resolved by positioning data from a Bluetooth beacon \cite{Tommy2016}.

Our present goal is to study the opportunity for similar approaches in the domestic Ubicomp context. Our specific interest is that, unlike the controlled conditions in which the positioning accuracy of Bluetooth beacons is normally calibrated \cite{Faragher2015}, private houses contain a number of unpredictable elements that are known to introduce challenges for the accuracy and reliability of Bluetooth positioning. These are discussed in more detail later, but include multi-path signal interference, variable surface reflections, attenuation due to human bodies and so on. So many factors affect the accuracy of these location technologies, in fact, that it would be extremely challenging to measure and calibrate them all – even for an individual house, let alone to create a generic model that can be transplanted to any house.

Instead, our approach inspired by Dourish and Chalmers is to treat the house itself as an ``ethnographic'' object. We do not mean in the sense that we study people’s behaviour in their houses (although that will come as a later stage in our research). Rather, we follow the example of ethnographic design theorists such as urban planner Kevin Lynch \cite{Lynch1960} and architect Christopher Alexander \cite{Alexander1979}, studying the house itself as an ethnographic object that carries the human traces of its occupants. With this ethnographic intent, we have carried out surveys to understand what the near future of location sensing in the home might look like, and the extent to which it is a seamful resource for user interaction. We have drawn on a sample that is rich and diverse, rather than controlled, in order to offer an alternative to existing laboratory study techniques. In particular, we provide a user-oriented analysis of Bluetooth location in four very different homes, located in three countries.

\section{RELATED WORK}

The use of Wi-Fi and Bluetooth signals for indoor positioning is well established. The received signal strength from a radio transmitter decreases with distance from the source, but indoor spaces present complicated propagation environments, and so simple ranging models based on free space path loss are known to produce highly-variable indoor positioning performance \cite{Faragher2010}. This was demonstrated as early as 2000 by Microsoft when they compared these two methods \cite{Bahl2000}. Fingerprinting is now the standard approach employed by indoor location-based-service providers, but surveying schemes are required to log the locations of the fingerprints initially in order to later provide location-based services. The surveying problem can be solved by crowdsourcing \cite{Mazumdar2014} or by machine learning methods such as Simultaneous Localisation and Mapping \cite{Faragher2012, Ferris2007}.

Bluetooth Low Energy (BLE) \cite{Lindh2015, Heydon2012a, Miller2011} has been developed in order to provide a method of transmitting very short packets of information short distances in order to improve the efficiency of the Internet of Things. BLE beacons can also be used to provide indoor location based services through either proximity detection or fingerprinting. BLE and Wi-Fi both operate in the same 2.4 GHz radio band and so both signal types are affected by attenuation and antenna detuning caused by interactions with the human body. The channel bandwidth of BLE is also much smaller than for Wi-Fi channels and so the susceptibility of BLE to large signal strength variations due to multipath interference is much higher than for Wi-Fi \cite{Faragher2015}.

\section{STUDY DESIGN}

In the design of this ‘ethnographic’ study, our goal was to understand the ways in which Bluetooth signal strength could be interpreted in the technologically seamful environment of actual houses, rather than in the controlled and calibrated laboratory environment. This goal led to three key decisions with regard to the study design:
1. We wanted to understand context in a way that represented home-owner's ‘conversation’ with the technical functionality of their houses. As a result, we paid special attention to rooms in the house that had specific technical functions, rather than characteristics defined by spatial layout and infrastructure. In fact, conventional names for rooms in the home already reflect the validity of technical functionality in the semantic interpretation of spatial context - kitchens, bathrooms and laundries are all marked by their functional context, independent of other location cues.
2. Although one can imagine that future houses might have embedded location monitoring infrastructure, and indeed many ubicomp researchers are working to create such infrastructure, we wished to concentrate on the pragmatic and ‘seamful’ circumstances in which new technologies actually arrive in real houses. Although Bluetooth beacon capabilities may be embedded in a variety of devices, and even distributed around a house by technical enthusiasts wishing to engage in lifelogging or home automation, we decided to explore the more likely near-term scenario that this capability might first be deployed as an additional ‘IoT’ market feature in a new appliance – for example a refrigerator, washing machine, or shaving station (in the examples of functional spaces already described).
3. Rather than grid-based laboratory survey techniques, we wished to gain insight into the way that signal strength would be experienced by an actual resident in the house, using a commodity mobile device. We therefore designed survey routes that represented real walking paths through each home, and carried out the survey by walking along this path, holding the phone in the natural hand position of a standing user during interaction.
Apart from these ethnographic constraints that were chosen to represent seamful and functional context, all other aspects of the survey followed best practice in signal strength survey, as derived from our previous laboratory studies - we created an app that sampled signal strength at approximately 10cm intervals along the path, measured the length of the paths within an accurate floor plan, and repeated each walk several times, in order to assess variability.

\section{STUDY PROCEDURE}

We used a prototype low-energy Bluetooth beacon made by CSR. The beacon power setting was configured to transmit a signal capable of covering an open area of around 50 metres. We placed the beacon in different locations inside a house/apartment, and measured the signal strength at various locations using an Android smart phone, while carrying the phone, and walking along predefined paths in both directions three times. We repeated this procedure in two houses and two studio flats situated in different countries including England, Australia, and Sri Lanka. All the houses and apartments were inhabited at the time of the study and have different characteristics such as layout, ceiling height, and number of stories.

The English house (Figure 1) is a two-storey terraced building – a style of housing in which a row of identical houses share side walls. The English studio flat (Figure 2) is situated at ground level with no upper neighbours. It is technically part of a detached house – a house that does not share a wall with a neighbouring dwelling – and only shares one wall with the main house. The Australian studio flat (Figure 3) is on the first floor of a three floor building. There are 15 apartment units with similar layout on each floor. The house in Sri Lanka was two-stories (Figure 4), but with ceiling height approx. 4 meters, in contrast to approx. 2.4 meters for the English and Australian dwellings).

\begin{figure*}
  \centering
  \includegraphics[width=\textwidth]{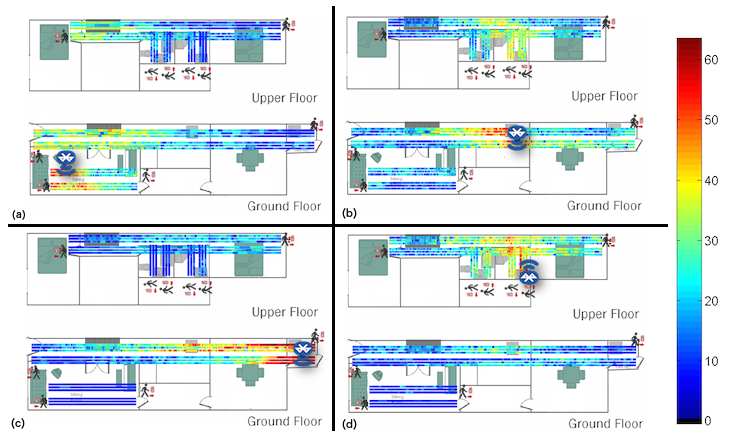}
 \caption{Signal strengths recorded in an English two-story terraced house. The beacon was placed in four different locations: (a) living room (b) kitchen (c) laundry room (d) bathroom. The colored lines indicate the walking paths and different colors represent different signal strengths (red highest and blue lowest). The direction of the beacon antenna is depicted using the direction of the waves in the Bluetooth icon.}
 \label{Figure:Settings}
\end{figure*}

We placed the beacon in functional locations such as kitchen, laundry room (if available), bathroom, bedroom, and living room. As discussed, the beacon was either placed on an electronic appliance such as refrigerator, washing machine or television, or attached to the ceiling with adhesive. The beacon antenna direction faced outward when on appliances (as shown in Figure 1). In ceiling locations, separate measurements were made with the antenna facing in each of four directions.
Data was collected using a commodity smartphone: a Nexus 4 running Android `KitKat' 4.4.4. We implemented a mobile app that collected all messages transmitted by the beacons and recorded their signal strengths for the entire duration of walking in the predefined paths. All survey measurements were made by the same person, walking at a steady natural speed, while holding the smart phone in a natural position. In the results reported below, the position of each signal strength reading is determined by linear interpolation along the path, based on timestamps of the beacon messages.

\section{RESULTS}

Typical results from the four different dwellings are explained with reference to Fig 1, which shows the floor plan of the English house. The observations reported below reflect the qualitative finding from all four properties. 
In Fig 1., the signal strength variations observed along each path are shown as a colour map. For each survey path, we collected data during six walks along the path – three times in each direction – to show both the variability in signals, and also the degree to which the natural walking pace results in consistency of time-interpolated positions.
The first observation is that there is considerable variation in the pattern of signal strength variation, depending on the direction in which the user walks. This is due to attenuation of the signal by the user’s body, when facing away from the beacon. In Fig 1c), the signal level in the next room when walking away from the beacon is the same as that three rooms away when walking toward it. This effect is occasionally reversed in the same room where the beacon is placed, apparently as a result of reflection from an opposite wall.
The second observation is that in this (brick) two-story house, the wooden floor between stories is rather permeable to the signal. As a result, In Fig 1a) the signal from a beacon on the TV set is stronger in the hallway upstairs than it is at the other end of the room where the TV is located. This effect was also noted in the Sri Lankan house. Of course this is not a problem in the two single-floor apartments, although other effects (reflections and doorways) became more salient in those smaller dwellings. In general, larger houses provided better support for separation of functional locations.
Partition walls allow Bluetooth signals to pass easily, with less attenuation than presented by the user’s body. This was a major factor in the small apartments, and can be seen between the two upstairs bathrooms in Fig 1d). This could present a significant obstacle to the type of ‘conversational’ context setting that we had envisioned, in that the apparent context of a functional space may be completely different in the room next door.
A further seamful consequence observed in signal propagation is that signal strength is relatively high when passing the open doorway of a room containing a beacon, especially when there is a line of sight to the beacon location. This is seen in Fig 1a), and especially markedly in Fig 1d), where walking along the hall presents the same signal strength as in the room containing a beacon. It would be difficult for a user immediately to diagnose this cause in Fig 1a), because the apparent signal path comes from a beacon location that is not visible through the door – apparently having been reflected through the door from a metal fireplace screen on the far wall.

\section{IMPLICATIONS FOR DESIGN}

We have presented a brief summary of signal strength measurements in a naturalistic situation, in order to show the ways in which location technologies do not (yet) support the functional ‘conversations’ that are essential to contextual interaction in Ubicomp.

Existing applications of Bluetooth beacons typically expect that the user is standing in close proximity facing the beacon, in order to avoid the ambiguity that results from multiple paths and body attenuation. In our previous work, we have explored seamful experiences designed around the observation that, although we might not know where the user is, we are reasonably confident that he is not near the beacon \cite{Tommy2016}.

We have described our observational approach as ‘ethnographic’, in order to contrast it with the calibrated laboratory measurements of signal strength-based location sensing that we have carried out in the past. However, even in the course of this study, it was clear that ‘expert’ usage of the signal strength measurement device (a conventional mobile phone) was essential to obtaining reliable results. Our earliest surveys resulted in contradictory and inconsistent readings far beyond those shown in Fig 1. More consistent results as our project continued represent a kind of ‘taming’ of the measurements intended to be made ‘in the wild’. This can be compared to the well-known finding from laboratory studies, that replicability of experimental results depends on the social context in which the work is done [14]. It is interesting to speculate how far this kind of calibration work might be necessary in order for householders to work with context in domestic settings.
In the first dwelling we surveyed (the studio flat in Australia), we compared ceiling-mounted beacons to beacons embedded in appliances. This scenario more closely resembles the current market expectation for location beacons, which are often sold in a stick-on package so that they can be deployed as location infrastructure. However, despite apparently unambiguous positioning (the centre of a ceiling in a small room), these free-standing beacons were even more ambiguous than opportunistic placement in appliances, because they allowed a greater range of reflections, signal paths through doors and so on. This appears to be an important piece of design guidance for determining functional context, given that so many functional appliances are explicitly linked to the functional rooms in which they are found.

\section{Conclusion}
Location sensing is a key enabling technology in order for Ubicomp to support contextually-informed interaction. Most calibrated testing of location-sensing devices takes place in the controlled environment of laboratories. However laboratories are very different to the domestic situations in which `context' has been identified as a problematic social construct. In this study, we have taken a systematic but contextually-informed approach to the use of Bluetooth signal strength as a location sensing technique. We have made systematic measurement surveys, informed by laboratory studies, but in a diverse range of domestic settings. The detailed design of these surveys has been motivated by the natural environment implied in the Bluetooth beacon standards – relating the technical situation of the beacon to the functional semantics of different spaces within the home. This research method can be considered as a situated, ‘ethnographic’ response to the study of the physical infrastructure in houses, as opposed to their occupants, whose lives are reflected by that infrastructure. The results offer insights to the future design of `seamful' approaches to indoor location sensing, and to the ways that context might be constructed and interpreted in a seamful manner.

\section*{Acknowledgement}

This work is supported by a Swiss National Science Foundation Early Postdoc Mobility fellowship (P2TIP2 152264), and is also partially funded by International Alliance of Research Universities (IARU) Travel Grant and The ANU Vice
Chancellor Travel Grant.

\bibliographystyle{abbrv}
\bibliography{library}

\begin{thebibliography}{10}

\bibitem{Alexander1979}
C.~Alexander.
\newblock {The Timeless Way of Building}, 1979.

\bibitem{Bahl2000}
P.~Bahl and V.~N. Padmanabhan.
\newblock {RADAR: an in-building RF-based user location and tracking system}.
\newblock In {\em Proceedings IEEE INFOCOM 2000. Conference on Computer
  Communications. Nineteenth Annual Joint Conference of the IEEE Computer and
  Communications Societies (Cat. No.00CH37064)}, volume~2, pages 775--784
  vol.2, 2000.

\bibitem{Chalmers2003}
M.~Chalmers and I.~MacColl.
\newblock {Seamful and Seamless Design in Ubiquitous Computing}.
\newblock In {\em Workshop At the Crossroads: The Interaction of HCI and
  Systems Issues in UbiComp.}, number January, page~8, 2003.

\bibitem{Dourish2004}
P.~Dourish.
\newblock {What we talk about when we talk about context}.
\newblock {\em Personal and Ubiquitous Computing}, 8(1):19--30, 2004.

\bibitem{Faragher2015}
R.~Faragher and R.~Harle.
\newblock {An Analysis of the Accuracy of Bluetooth Low Energy for Indoor
  Positioning Applications}.
\newblock Technical report, University of Cambridge, 2015.

\bibitem{Faragher2010}
R.~M. Faragher and P.~J. Duffett-Smith.
\newblock {Measurements of the effects of multipath interference on timing
  accuracy in a cellular radio positioning system}.
\newblock {\em IET Radar, Sonar Navigation}, 4(6):818--824, dec 2010.

\bibitem{Faragher2012}
R.~M. Faragher, C.~Sarno, and M.~Newman.
\newblock {Opportunistic radio SLAM for indoor navigation using smartphone
  sensors}.
\newblock In {\em Record - IEEE PLANS, Position Location and Navigation
  Symposium}, pages 120--128, 2012.

\bibitem{Ferris2007}
B.~D. Ferris, D.~Fox, and N.~Lawrence.
\newblock {WiFi-SLAM using Gaussian process latent variable models}.
\newblock {\em Science}, pages 2480--2485, 2007.

\bibitem{Lindh2015}
J.~Lindh.
\newblock {Bluetooth Low Energy Beacons}.
\newblock {\em Texas Instruments}, (January):1--13, 2015.

\bibitem{Lynch1960}
K.~Lynch.
\newblock {\em {The Image of the City}}.
\newblock 1960.

\bibitem{Mazumdar2014}
P.~Mazumdar, V.~J. Ribeiro, and S.~Tewari.
\newblock {Generating indoor maps by crowdsourcing positioning data from
  smartphones}.
\newblock In {\em 2014 International Conference on Indoor Positioning and
  Indoor Navigation (IPIN)}, pages 322--331, oct 2014.

\bibitem{Tommy2016}
T.~Nilsson, C.~Hogsden, C.~Perera, S.~Aghaee, D.~Scruton, A.~Lund, and A.~F.
  Blackwell.
\newblock {Applying Seamful Design in Location-based Mobile Museum
  Applications}.
\newblock {\em ACM Transactions on Multimedia Computing Communications and
  Applications (TOMM)}, 12(4):56:1----56:23, 2016.

\end{thebibliography}

\newpage
\onecolumn
\appendix[UK Studio Flat]

\begin{figure*}[h!]
  \centering
  \includegraphics[scale = 0.60]{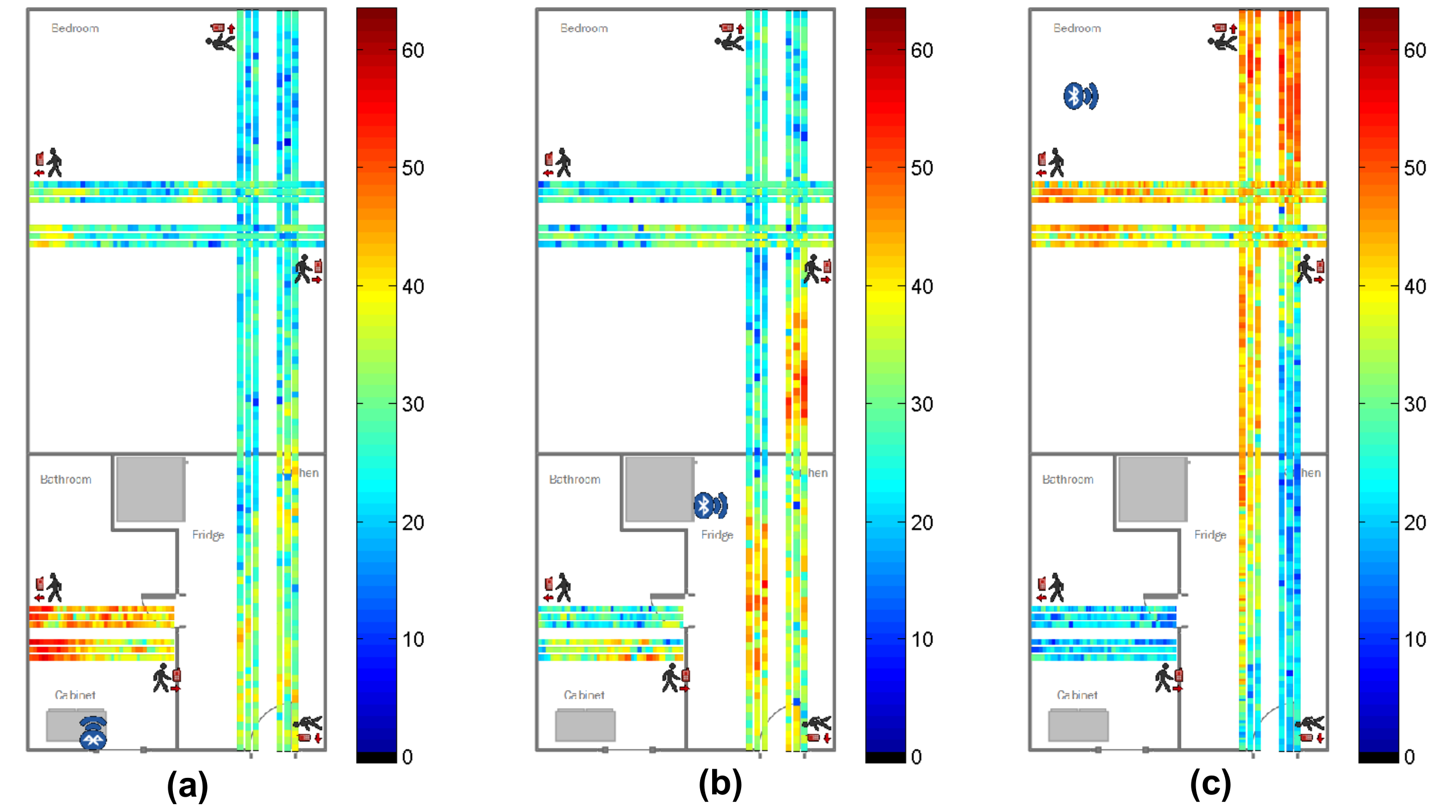}
 \caption{Signal strengths recorded in an UK Studio flat. The beacon was placed in three different locations: (a) bathroom (on a cabinet) (b) living room (c) kitchen (on the fridge). The colored lines indicate the walking paths and different colors represent different signal strengths (red highest and blue lowest). The direction of the beacon antenna is depicted using the direction of the waves in the Bluetooth icon.}
 \label{Figure:UKStudio}
\end{figure*}


\begin{figure*}[h!]
  \centering
  \includegraphics[scale = 0.34]{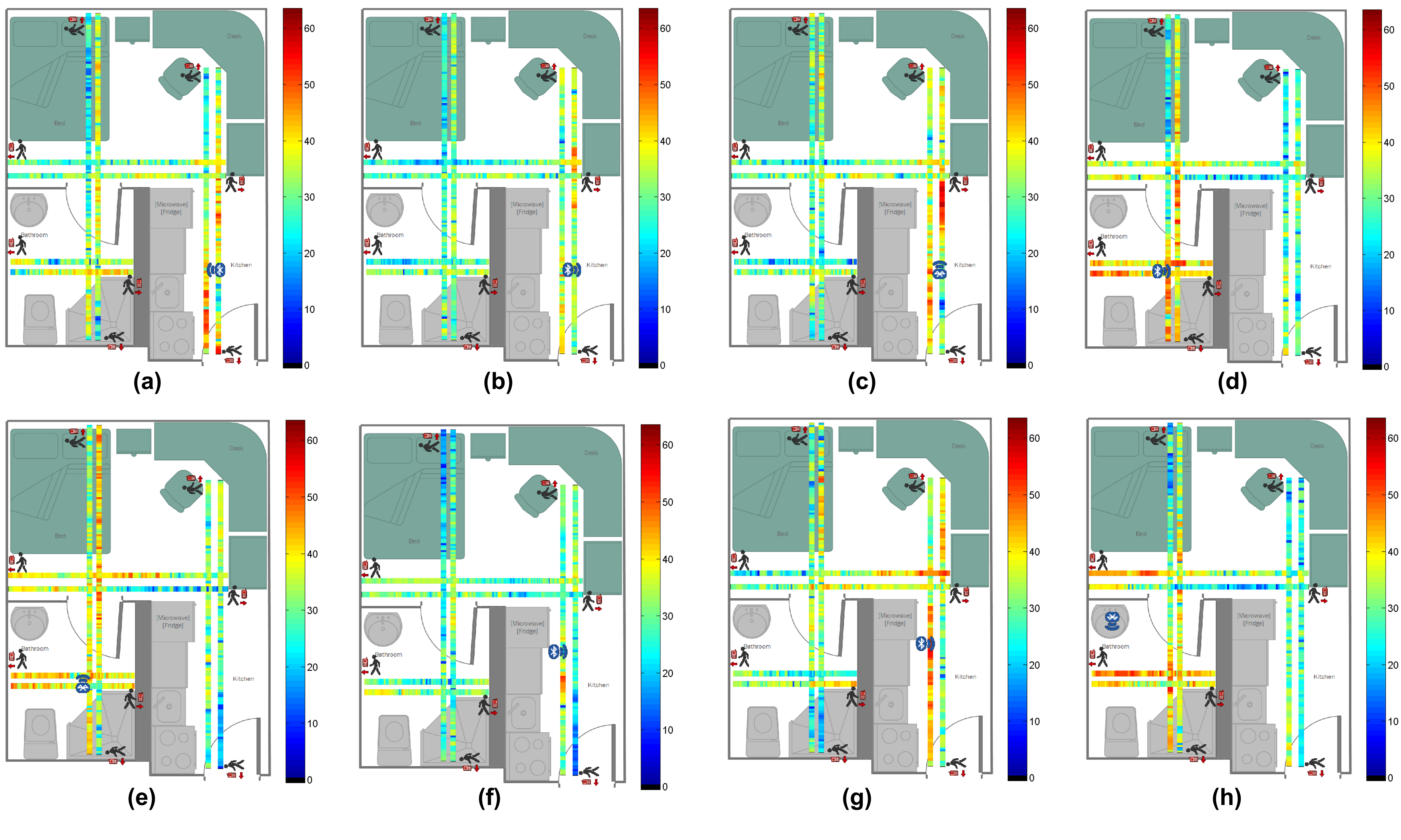}
 \caption{Signal strengths recorded in an Australian Studio flat. The beacon was placed in eight different locations: (a) kitchen ceiling (antenna towards east), (b) kitchen ceiling (antenna towards west), (c) kitchen ceiling (antenna towards north), (d) bathroom celling (antenna towards west), (e) bathroom celling (antenna towards north), (f) inside the bar fridge in kitchen, (g) outside the bar fridge in kitchen, (h) on the sink in the bathroom. The colored lines indicate the walking paths and different colors represent different signal strengths (red highest and blue lowest). The direction of the beacon antenna is depicted using the direction of the waves in the Bluetooth icon.}
 \label{Figure:AUSStudio}
\end{figure*}


\begin{figure*}[h!]
  \centering
  \includegraphics[scale = 0.44]{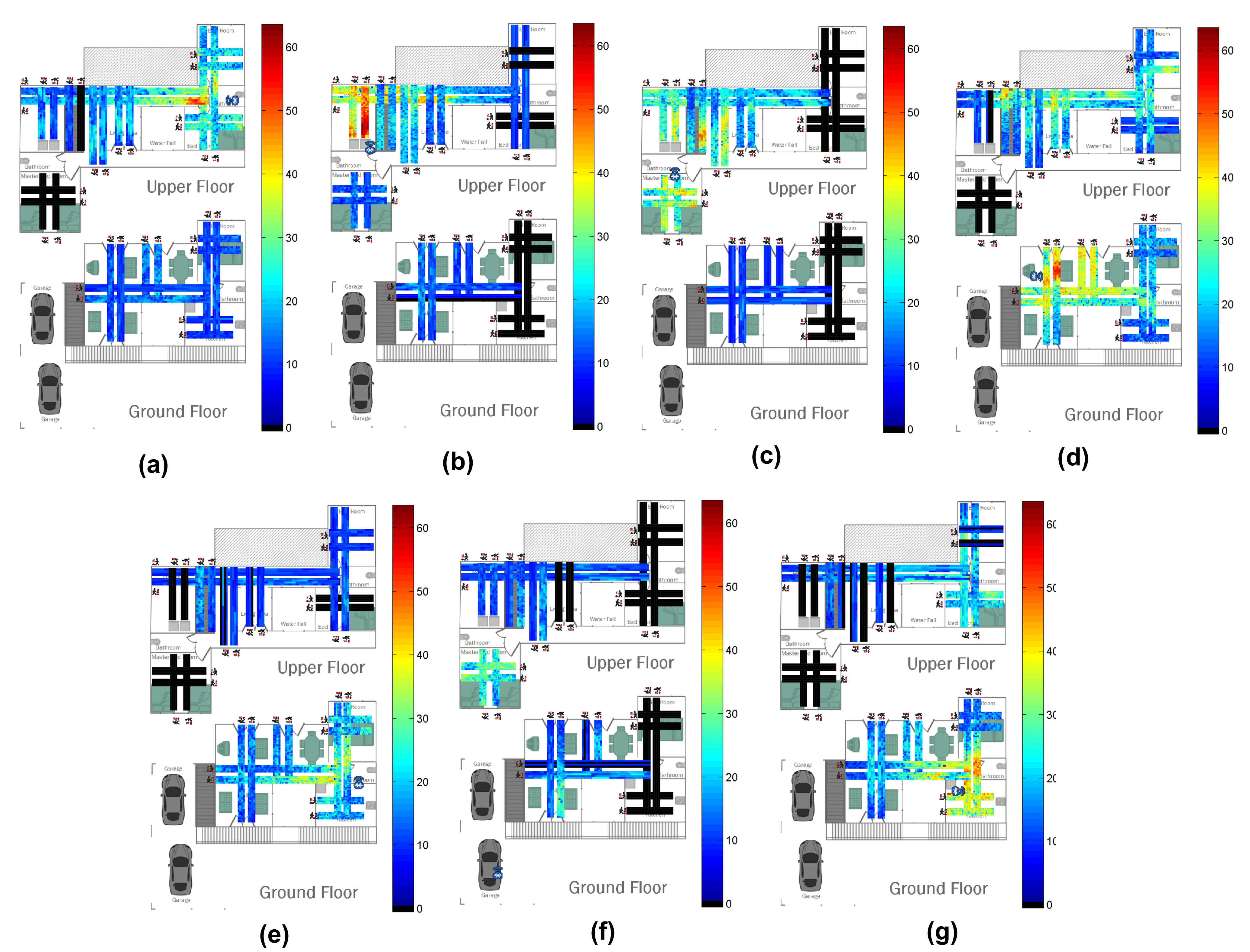}
 \caption{Signal strengths recorded in a two story house in Sri Lanka. The beacon was placed in seven different locations: (a) upper floor bathroom, (b) upper floor laundry room, (c) upper floor master bedroom bathroom, (d) ground floor living room on top of TV, (e) ground floor bathroom, (f) in the garage on the the car, (g) on the fridge in the kitchen.  The colored lines indicate the walking paths and different colors represent different signal strengths (red highest and blue lowest). The direction of the beacon antenna is depicted using the direction of the waves in the Bluetooth icon.}
 \label{Figure:SLHouse}
\end{figure*}

%

%


\ifCLASSOPTIONcaptionsoff
  \newpage
\fi

\end{document}